\documentclass[pre,twocolumn,amsmath,amsfonts,amssymb,showpacs]{revtex4} 
\usepackage{graphicx}
\usepackage{mathptmx}      
\usepackage{latexsym}
\def\beq{\begin{equation}}
\def\eeq{\end{equation}}
\def\bea{\begin{eqnarray}}
\def\eea{\end{eqnarray}}

\begin{document}

\title{Generalized relation between the relative entropy and
dissipation for nonequilibrium systems }

\author{Pegah Zolfaghari}

\author{Somayeh Zare}

\author{Behrouz Mirza}
\email{b.mirza@cc.iut.ac.ir}
\affiliation{Department of Physics, Isfahan University of Technology, Isfahan 84156-83111, Iran}



\begin{abstract}
Recently, Kawai, Parrondo, and Van den Broeck have related
dissipation to time-reversal asymmetry. We generalized the result
by considering a  protocol where the physical system is driven
away from an initial thermal equilibrium state with temperature
$\beta_0$  to a final thermal equilibrium state at a different
temperature. We illustrate the result using a model with an exact
solution, i.e., a particle in a moving one-dimensional harmonic
well.

\end{abstract}
\maketitle

\section{Introduction}

Irreversible thermodynamic processes are the ones that cannot be
closed. In other words, the system and its surroundings never
return to their original states. There are a number of features
associated with such processes which include (i) dissipation; (ii)
asymmetry in the arrow of time; and (iii) broken equilibrium. In
recent years, some relations have been stated between dissipation
and time-reversal asymmetry \cite{0,01,02,03,3,q,2,exper1,
exper2, exper3}. Here, we will focus on the relation obtained by
Kawai, Parrondo, and Van den Broeck (KPV), which is expressed by
the following \cite{3}, \bea \label{1}\beta\langle
W_{diss}\rangle\geq
D[\rho_{F}(z,t)\parallel\rho_{R}(z^{\ast},\tau-t)]. \eea In this
relation, $\langle W_{diss}\rangle$ is the average  work
dissipated during the process in which the system evolved from one
canonical equilibrium state at a temperature $T$  into another at
the same temperature. Based on the second law, the average work
performed on the system must exceed the difference between  the
free energy in the initial equilibrium state and that in the final
equilibrium one, i.e. $\langle W\rangle \geq \Delta F=
F_{B}-F_{A}$ \cite{4}. The dissipated work is defined as  $\langle
W_{diss}\rangle=\langle W\rangle - \Delta F$.
$D(\rho_{F}(z,t)\parallel\rho_{R}(z^{\ast},\tau-t))$  denotes the
relative entropy \cite{5}, a measure of the distinction between
$\rho_{F}$, i.e. the time-dependent phase-space density, during
the  forward process ($A\rightarrow B$) and $\rho_{R}$, i.e. the
time-dependent phase-space one,
 during the reverse process ($B\rightarrow A$). $z=({\bf x},{\bf p})$
designates a point in the phase space, and the asterisk denotes
the reversal of momenta, ${\bf p}\longrightarrow-{\bf p}$. For the
case of the Hamiltonian dynamics, where the system evolves
deterministically, Eq. (1) is an equality; i.e., the average
dissipated work can be expressed by \cite{2}, \bea \label{kawai2}
\nonumber\beta\langle W_{diss}\rangle &=& \int dz \rho_{F}(z,t)
\ln[\frac{\rho_{F}(z,t)}{\rho_{R}(z^{\ast},\tau-t)}]\\&=&
D[\rho_{F}(z,t)\parallel\rho_{R}(z^{\ast},\tau-t)]. \eea

 Consider a system that is driven far from equilibrium by a protocol, where
the inverse temperature of the initial and final equilibrium
states are $\beta_{0}$ and  $\beta_{\tau}$, respectively. Our goal
is to generalize  Eq. (2) to this kind of process. A
 basic motivation to study this generalized relation is its possible
application to the evolution of black holes that we will discuss
elsewhere. This will be attempted in Sec. 2. In Sec. 3, we will
consider two solvable models for the generalized equation.
\section{Generalization of  KPV equation}
We consider a system that is initially coupled with a reservoir
in the  reverse temperature $\beta_{0}$.  The reservoir is then
removed. Subsequently, from $t=0$ to a later time $t=\tau$, the
external forces are turned on according to some arbitrary but
predetermined schedule, or protocol, $\lambda_{t}$. The
microscopic evolution of the system during this time interval is
described by the trajectory  $z_{t}=({\bf p}, {\bf q})$. The
Hamiltonian of the system is denoted by $H(\lambda_{t}, z_{t})$.
At time $t=\tau$, the system is coupled with a reservoir in the
reverse temperature $\beta_{\tau}$, again. Similar to the
derivation of the Crooks equality \cite{2}, we consider two
processes which are labeled forward $(F)$ and reverse $(R)$.
 The initial phase-space densities for the forward and reverse processes are given by $(I \equiv F\  $or$ \ R)$
 \bea
 \label{2}
\rho_{0}^{I}(z_{0},\lambda_{0})=\frac{1}{Z_{0}}\exp[-\beta_{0}H(z_{0}^{I},\lambda_{0})],
\eea

\noindent  and $ Z_{t}=\int dz_{t}\exp[-\beta_{t} H(z_{t},
 \lambda_{t})]   $ is  partition function at time $t=0 $
 or $t=\tau$. It should be noted that temperature is only defined for
the initial and final states and not for any state between the
two. By combining Eq. (\ref{2}) and definition of the partition
function, we get \bea \label{5}
\frac{\rho_{0}^{F}}{\rho_{0}^{R}}=\frac{Z_{\tau}}{Z_{0}}\exp[\beta_{\tau}
H(z^{R}_{0}; \lambda_{\tau})-\beta_{0}H(z^{F}_{0}; \lambda_{0})].
\eea
 According to the definition of free energy given by  $ F(\lambda_{t})=-\beta_{t}^{-1}\ln
 Z(\lambda_{t})$,
 Eq. (\ref{5}) can be rewritten as follows:
\bea \label{6}
 \frac{\rho_{0}^{F}}{\rho_{0}^{R}}=\exp[-\Delta(\beta F)]\exp[\beta_{\tau} H(z^{R}_{0}; \lambda_{\tau})-\beta_{0}H(z^{F}_{0};
 \lambda_{0})].
 \eea
 Note that  $\Delta(\beta F)= \beta_{\tau}F_{\tau} - \beta_{0}F_{0}$. Assume that the work performed on the system is defined by  \cite{2,work1,work2,
 classic
system}
 \bea
\label{7}
 W[z_{t}]&=&\int_{0}^{\tau} dt
\dot{\lambda} \frac{\partial H}{\partial
\lambda}(z_{t},\lambda_{t}).
 \eea
Then, based  on this definition, we have \cite{2}
 \bea
\label{8}
 W[z_{t}]
 =H(z_{\tau}^{F};\lambda_{\tau})-H(z_{0}^{F};\lambda_{0}).
\eea Using Eq. (\ref{8}), Eq. (\ref{6}) can be simplified as  \bea
\label{9}
 \frac{\rho_{0}^{F}}{\rho_{0}^{R}}=\exp[-\Delta(\beta F)+\beta_{\tau}W+
 (\beta_{\tau}-\beta_{0})H_{0}],
 \eea
 where  $H_{0}\equiv H(z_{0}^{F};\lambda_{0})$. We can rewrite Eq. (\ref{9}) as
 \bea
\label{10} \ln (\frac{\rho_{0}^{F}}{\rho_{0}^{R}})=-\Delta(\beta
F)+\beta_{\tau}W+ (\beta_{\tau}-\beta_{0})H_{0}. \eea

\noindent We can use the definition of average as, $\int dz_{0}
\rho_{0}(z_{0}) A = <A>_{\rho_{0}}$, where, $A$ is an arbitrary
normalized function [$\int dz_{0} A(z_{0}, 0)=1$]. Therefore, we
can write equality (\ref{10}) as  
\bea \label{11} 
&&\int dz_0\ln
(\frac{\rho_{0}^{F}}{\rho_{0}^{R}})\rho_{0}^{F}=\langle\ln
(\frac{\rho_{0}^{F}}{\rho_{0}^{R}})
\rangle_{\rho_{0}^{F}}=\nonumber\\
&&-\Delta(\beta F)+\beta_{\tau}\langle W
\rangle_{\rho_{0}^{F}} + (\beta_{\tau}-\beta_{0})\langle
H_{0}\rangle_{\rho_{0}^{F}}.\eea Since the phase-space density is
conserved along any Hamiltonian trajectory, i.e.,
$\rho_{0}^{F}=\rho^{F}$ and $\rho_{0}^{R}=\rho^{R}$ and based on
the definition of the relative entropy 
 \bea
 \int dz \rho_{F}(z,t)
 \ln[\frac{\rho_{F}(z,t)}{\rho_{R}(z^{\ast},\tau-t)}]=
 D[\rho_{F}(z,t)\parallel\rho_{R}(z^{\ast},\tau-t)],\nonumber\\
 \eea 
we obtain
the following generalized form of Eq. (\ref{kawai2}):
\bea\label{111} D(\rho^{F}\parallel\rho^{R})=-\Delta(\beta
F)+\beta_{\tau}\langle W \rangle_{\rho_{0}^{F}} +
(\beta_{\tau}-\beta_{0})\langle H_{0}\rangle_{\rho_{0}^{F}}. \eea

The above expression is valid for the deterministic trajectories
of the system, including information about every degree of
freedom. If only partial information about the system is
available, the relative entropy is reduced and we will have an
inequality.\\ Since $z=(q,p)$ in deterministic dynamics
represents all position and momentum variables and, $\rho(q,p)$
surrounds the phase-space trajectory going through $z=(q,p)$,
Therefore, we can consider a partition of the entire phase space
\cite{3}.This partition is described by a sequence $z_{0},
z_{1},...z_{n}$. Corresponding  phase space distributions for the
forward and backward processes are given by  \bea \label{12}
\rho_{n}^{F}=\int_{z_{n}}\rho_{0}^{F}(p,q) dqdp, \qquad
 \rho_{n}^{R}=\int_{z_{n}^{*}} \rho_{0}^{R}(p,q) dqdp.
 \eea By integrating Eq. (\ref{10}) over $z_{n}$ we obtain
  \bea
 \label{13}
&&\langle \exp[-\beta_{\tau} W-(\beta_{\tau}-\beta_{0})H_0]\rangle_{n}^{F}=\nonumber\\&&\frac{\int_{z_{n}} \rho_{0}^{F}(p,q) \exp(-\beta_{\tau} W -(\beta_{\tau}-\beta_{0}) H_{0})dqdp} {\rho_{n}^{F}}\nonumber\\
 &=&\frac{\rho_{n}^{R}}{\rho_{n}^{F}}\exp[-\Delta(\beta
F)].
 \eea
 Now, by using  $\langle\exp(-x)\rangle\geq\exp\langle-x\rangle $, we can
rewrite Eq. (\ref{13}) as: \bea \label{15} \langle \beta_{\tau}
W+(\beta_{\tau}-\beta_{0})H_{0}\rangle_{n}^{F}\geq \Delta(\beta
F)+\ln\frac{\rho_{n}^{F}}{\rho_{n}^{R}}. \eea By performing an
average over the different subsets, we will have  
\bea \label{16}
\langle \beta_{\tau}
W+(\beta_{\tau}-\beta_{0})H_{0}\rangle^{F}&=&\Sigma_{n}\rho_{n}^{F}\langle
\beta_{\tau} W+(\beta_{\tau}-\beta_{0})H_{0}\rangle_{n}^{F}. \nonumber \\
\eea
Finally, considering Eqs. (\ref{15}) and (\ref{16})  yields
 the following generalized relations:\bea
\langle \beta_{\tau}W+(\beta_{\tau}-\beta_{0})H_{0}\rangle^{F}- \Delta(\beta F)&\geq&\int \ln(\frac{\rho_{n}^{F}}{\rho_{n}^{R}})d\rho_{n}^{F},\\
\label{24}\langle \beta_{\tau}
W+(\beta_{\tau}-\beta_{0})H_{0}\rangle^{F}-\Delta(\beta F)&\geq&
D({\rho_{n}^{F}}\parallel \rho_{n}^{R}). \eea





\noindent Equation (\ref{24}) is the basic result of this Brief
Report. In Sec. 3, we will consider an example that corresponds to
an exact equality as in Eq. (\ref{111}).
\section{Particle in a moving harmonic well }
In this section, we analyze Eq. (\ref{111}) for a particle with
mass $m$ which is trapped in a harmonic well with a spring
constant $k$. The Hamiltonian of the particle is given by

 \bea
\label{exam}H(x,p,\lambda)=\frac{p^2}{2m}+\frac{k}{2}(x-\lambda)^2.
\eea We will consider processes during which the center of the
well is moved either rightward or leftward at a constant speed,
$u$. These correspond to the forward and reverse protocols,
$\lambda_{F}(t)=ut$ and $\lambda_{R}(t)=u(\tau-t)$, where $\tau$
is the total time interval. Along this time interval, the initial
and final  reverse temperatures are   $\beta_{0}$ and
$\beta_{\tau}$, respectively,  during the forward process, and
vice versa  during the reverse process. We consider two dynamics
for this example: Hamiltonian dynamics and Langevin dynamics.

\subsection{Hamiltonian dynamics }
In this section we assume that the system is thermally isolated
from environment after the initial equilibration stage. Explicit
expressions for the initial equilibrium densities are given by the
following Gaussian distributions: \bea
\label{gf}\rho_{F}(z,0)&=&\frac{\beta_{0}}{2\pi}\sqrt{\frac{k}{m}}\exp[-\beta_{0}(\frac{p^2}{2m}+\frac{kx^2}{2})],
\\
\label{gr}\rho_{R}(z,0)&=&\frac{\beta_{\tau}}{2\pi}\sqrt{\frac{k}{m}}\exp[-\beta_{\tau}(\frac{p^2}{2m}+\frac{k}{2}(x-u\tau)^2)].
\eea
 By considering that in this situation the system evolves under the Hamiltonian dynamics, the equations of motions are given by
\bea \label{eqh}\dot{x_{F}}&=&\frac {p_{F}}{m},\qquad
\dot{p_{F}}=-k(x_{F}-ut),\\
\label{eqh1}\dot{x_{R}}&=&\frac {p_{R}}{m},\qquad
\dot{p_{R}}=-k(x_{R}-u\tau+ut). \eea Due to the Gaussian nature of
the initial densities [Eqs. (\ref{gf}) and (\ref{gr})] and the
linearity of the  equations of motion, the distribution remains
Gaussian for all times. Therefore, according to the relations of
means and covariances for two-dimensional Gaussian distributions,
($f_{G}(z)=\frac{1}{2\pi\sqrt{det\sigma}}\exp[\frac{1}{2}(z-\overline{z})^{T}.\sigma^{-1}.(z-\overline{z})]$),
\cite{2}
 and  consider that the
relative entropy between two Gaussian distributions, $f_{G}(z)$
and $g_{G}( z^{\ast})$, is \bea \label{e}\nonumber
D[f_{G}(z)\parallel
g_{G}(z^{\ast})]&=&-1+\frac{1}{2}[\ln(\frac{\det\sigma_{g}}{\det
\sigma_{f}})+Tr(\sigma^{-1}_{g}.\sigma_{f}^{\ast})]\\&+&\frac{1}{2}(\overline
z_{f}^{\ast}-\overline z_{g})^{T}. \sigma_{g}^{-1}.(\overline
z_{f}^{\ast}-\overline z_{g}), \eea
 where  $\sigma^{\ast}_{xp}=-\sigma_{xp}$ and all other elements of $\sigma^{\ast}$
 are unaltered \cite{sigma}. 
 We have  \bea
\label{right}\nonumber D[\rho_{F}(z,t)\parallel\rho_{R}(z^{\ast},\tau-t)]&=&-1+\ln(\frac{\beta_{0}}{\beta_{\tau}})+(\frac{\beta_{\tau}}{\beta_{0}})\\
&+&\beta_{\tau}mu^{2}[1-\cos(w\tau)]. \eea By considering  the
relations for means and variances that we obtain here, we can say
that the  means and variances are the same with the results in
Ref. \cite{2}; the only difference is  that here  we have
$\beta_{0}$ in $t=0$  and $\beta_{\tau}$ in $t=\tau$.
 Also, according to the following relations
 \bea
 \label{ww}\nonumber\langle W\rangle_{\rho_{0}^{F}}&=&
\nonumber-uk\int_{0}^{\tau}[\overline{x}_{F}(t)-ut]\\&=&uk\int_{0}^{\tau}
\nonumber[\frac{\sin(wt)}{w}]\\&=& mu^2[1-\cos(w\tau)],
 \eea
  \bea
  \label{h0}\langle
H_{0}\rangle_{\rho_{0}^{F}}=\frac{1}{\beta_{0}},\qquad
-[\Delta(\beta F)]=\ln(\frac{\beta_{0}}{\beta_{\tau}}). \eea The
left-hand side of Eq. (\ref{111}) is equal to \bea
\label{n}\beta_{\tau}mu^2[1-\cos(w\tau)]+(\beta_{\tau}-\beta_{0})\frac{1}{\beta_{0}}+\ln(\frac{\beta_{0}}{\beta_{\tau}}).
\eea So, we have 
 \bea \label {m}
 &&\beta_{\tau}\langle
 W\rangle_{\rho_{o}^{F}}+(\beta_{\tau}-\beta_{0})\langle
 H_{0}\rangle_{\rho_{0}^{F}}-\Delta(\beta F)
 =\nonumber\\
 &&(\beta_{\tau}-\beta_{0})\frac{1}{\beta_{0}}+\ln(\frac{\beta_{0}}{\beta_{\tau}})+\beta_{\tau}mu^2[1-\cos(w\tau)].
 \eea 
 Therefore, comparison of  Eqs. (\ref{right}) and (\ref{m})
yields the following equation for  a Hamiltonian dynamics:

\bea
&&\beta_{\tau}{\langle
W\rangle}_{\rho_{0}^{F}}+(\beta_{\tau}-\beta_{0})\langle
H_{0}\rangle_{\rho_{0}^{F}}-\Delta(\beta
F)=\nonumber\\
&&D[\rho_{F}(z,t)\parallel\rho_{R}(z^{\ast},\tau-t)], 
\eea
 which
is a special case of Eq. (\ref{111}).

\ \ \ 

\subsection{Overdamped Langevin dynamics }
In this section we consider a system that interacts with its
environment so that we may use overdamped Langevin
dynamics\cite{2}. For the forward process the Fokker-Planck
equation for $\rho_{F}(x,t)$ is \bea \frac{\partial}{\partial
t}\rho_{F}(x,t)=\frac{k}{\gamma}\frac{\partial}{\partial
x}[(x-ut)\rho_{F}(x,t)]+\frac{1}{\gamma\beta}\frac{\partial^{2}}{\partial
x^{2}}\rho_{F}(x,t). \eea where $\gamma$ is the friction
coefficient. By the use of the means and variances for forward and
reverse processes,  \bea \overline{x}_{F}(t)= ut-\frac{\gamma
u}{k}(1-e^{-kt/\gamma}),\ \ \
\sigma^{2}_{F}=\frac{1}{\beta_{0}k},\eea \bea
\overline{x}_{R}(t)= u(\tau-t)+\frac{\gamma
u}{k}(1-e^{-k(\tau-t)/\gamma}), \ \ \
\sigma^{2}_{R}=\frac{1}{\beta_{\tau}k}, \eea we have the following
results for the relative entropy and average work: 
\bea\label{40}
&&D[\rho_{F}(x,t)\parallel\rho_{R}(x,\tau-t)]=
-1+\ln(\frac{\beta_{0}}{\beta_{\tau}})+(\frac{\beta_{\tau}}{\beta_{0}})
\nonumber\\&&+\frac{2\beta_{\tau}\gamma^{2}u^{2}}{k}
\{1-e^{-k\tau/2\gamma}cosh[\frac{k}{\gamma}(\frac{\tau}{2}-t)]\}^{2},
\eea \bea
 \nonumber\langle W\rangle_{\rho_{0}^{F}}&=&
\nonumber-uk\int_{0}^{\tau}[\overline{x}_{F}(t)-ut]\\&=&\gamma
u^{2}[\tau-\frac{\gamma }{k}(1-e^{-k\tau/\gamma})].\eea In this
situation, we have to demonstrate the validity of inequality
(\ref{24}). We note that the left-hand side of this inequality is
\begin{widetext}
\bea\label{42}\langle \beta_{\tau}
W+(\beta_{\tau}-\beta_{0})H_{0}\rangle^{F}-\Delta(\beta
F)=\beta_{\tau}\gamma u^{2}[\tau-\frac{\gamma
}{k}(1-e^{-k\tau/\gamma})]+(\beta_{\tau}-\beta_{0})\frac{1}{\beta_{0}}+\ln(\frac{\beta_{0}}{\beta_{\tau}}),
\eea 
\end{widetext}
 and $D({\rho_{n}^{F}}\parallel
\rho_{n}^{R})$ has a  maximum value at $t={\tau / 2}$. Combining
Eqs. (\ref{40}) and (\ref{42})   we have
\bea\nonumber\langle\beta_{\tau}
W+(\beta_{\tau}-\beta_{0})H_{0}\rangle^{F}-\Delta(\beta F)
 &\geq& D({\rho_{n}^{F}}\parallel
\rho_{n}^{R}),\eea where the inequality is valid for any $\zeta
=\frac{k\tau}{\gamma} \geq 0$. Therefore the validity of
inequality (\ref{24}) for overdamped Langevin dynamics is
demonstrated.
\section{Conclusion }

We extended a known relation between the dissipated work and
time-reversal asymmetry to a more general case by assuming a
protocol which starts  from an equilibrium state and moves to
another equilibrium where the initial and final temperatures are
different (it should be noted that we do not need to define
temperatures between the two equilibria). Time-reversal asymmetry
is more clear in this kind of process; however, the definition of
dissipated work is not completely understood in this situation.
It will be  interesting to consider more general cases, as they
will provide a better understanding of  the relationship between
the time-reversal asymmetry and dissipation and other aspects of
non-equilibrium statistical mechanics.


\end{document}